\documentclass[reprint,aps,prb,floatfix]{revtex4-2}
\usepackage{bm}
\usepackage{amssymb}
\usepackage{amsmath}
\usepackage{graphicx}
\usepackage{rotating}
\usepackage{epsfig}
\usepackage{psfrag}
\usepackage{subfigure}
\usepackage{braket}
\usepackage{tikz}
\usepackage{stmaryrd}
\usepackage{wasysym}
\usepackage{array}
\usepackage{booktabs}

\newcommand{\bea}{\begin{eqnarray}}
\newcommand{\eea}{\end{eqnarray}}
\newcommand{\bpm}{\begin{pmatrix}}
\newcommand{\epm}{\end{pmatrix}}

\usepackage[unicode=true,colorlinks=true,pdfpagemode=UseOutlines,pdfstartview=Fith]{hyperref}
\hypersetup{linkcolor=blue,citecolor=blue,urlcolor=blue}

\begin{document}
\title{Tuning entanglement phases and topological memory in the measurement-only Kitaev model with single and multi-qubit checks}

\author{Tushya Kalpada*, Aayush Vijayvargia*, Ezra Day-Roberts*, Onur Erten}
\affiliation{Department of Physics, Arizona State University, Tempe, AZ 85287, USA}

\begin{abstract}
Quantum circuits provide an emerging controllable platform to realize novel dynamical non-equilibrium phases including topologically ordered states. The Kitaev model has become a cornerstone of quantum magnetism due to its quantum spin liquid ground state and rich phase diagram. The Kitaev model has also been treated in the monitored circuit setting, giving rise to topological area-law and critical-law entanglement entropy phases. In this article, we study the evolution of its phase diagram under the addition of new terms, motivated by their effects in the Kitaev model.
We find that a single-qubit term, analogous to a magnetic field, leads to a trivial state in the high field limit, but with an additional intermediate volume-law phase. 
A three-qubit operator that commutes with the flux operators has the opposite effect: it stabilizes the critical-law phase against the short ranged area-law entanglement.
We also employ a four-qubit plaquette commuting operator that simultaneously measures two opposite identical-type bonds on a plaquette. This generates a distinct volume-law phase and preserves the plaquette fluxes and associated topological order, yielding extensive entanglement while coexisting with the topological memory characteristic of the area-law phase. We quantitatively locate phase boundaries using stabilizer (Clifford) simulations together with tripartite mutual information and entanglement entropy measures. Our results highlight the rich phase diagram accessible from the measurement-only Kitaev model as well as suggesting rules relating the newly added operators to the phases they promote.

\end{abstract}
\maketitle

\section{Introduction}
A central aim in nonequilibrium quantum physics is to pinpoint how interacting systems create entanglement, how it is structured in space and time, and under what conditions it can be stabilized. Monitored circuits, where local projective measurements replace or supplement unitary evolution, exhibit sharp steady state entanglement phases \cite{Skinner_PRX2019,Gullans_PRX2020,Choi_PRL2020,Fisher_annrev2023,Zabalo_PRL2022}. In measurement-only settings, the governing knob is not just the rate but the algebra and geometry of the measured operators: which operators are available, how they (fail to) commute, and how they are spatially arranged. Tuning these ingredients can realize area-law, volume-law, and critically (multiplicative logarithmic correction to area-law) entangled regimes \cite{Lavasani_NatPhys2021,Sang_PRR2021,Klocke_PRB2022,Ippoliti_PRX2021,Lavasani_PRL2021,Hastings_Quant2021,Vu_PRL2024}; symmetry and topology further enrich this landscape and can lead to phases with fault-tolerant quantum memory \cite{Agrawal_PRX2022,Barratt_PRL2022,Lavasani_NatPhys2021,Lavasani_PRL2021}.
\begin{figure}[t!]
    \centering
    \includegraphics[width=1\linewidth]{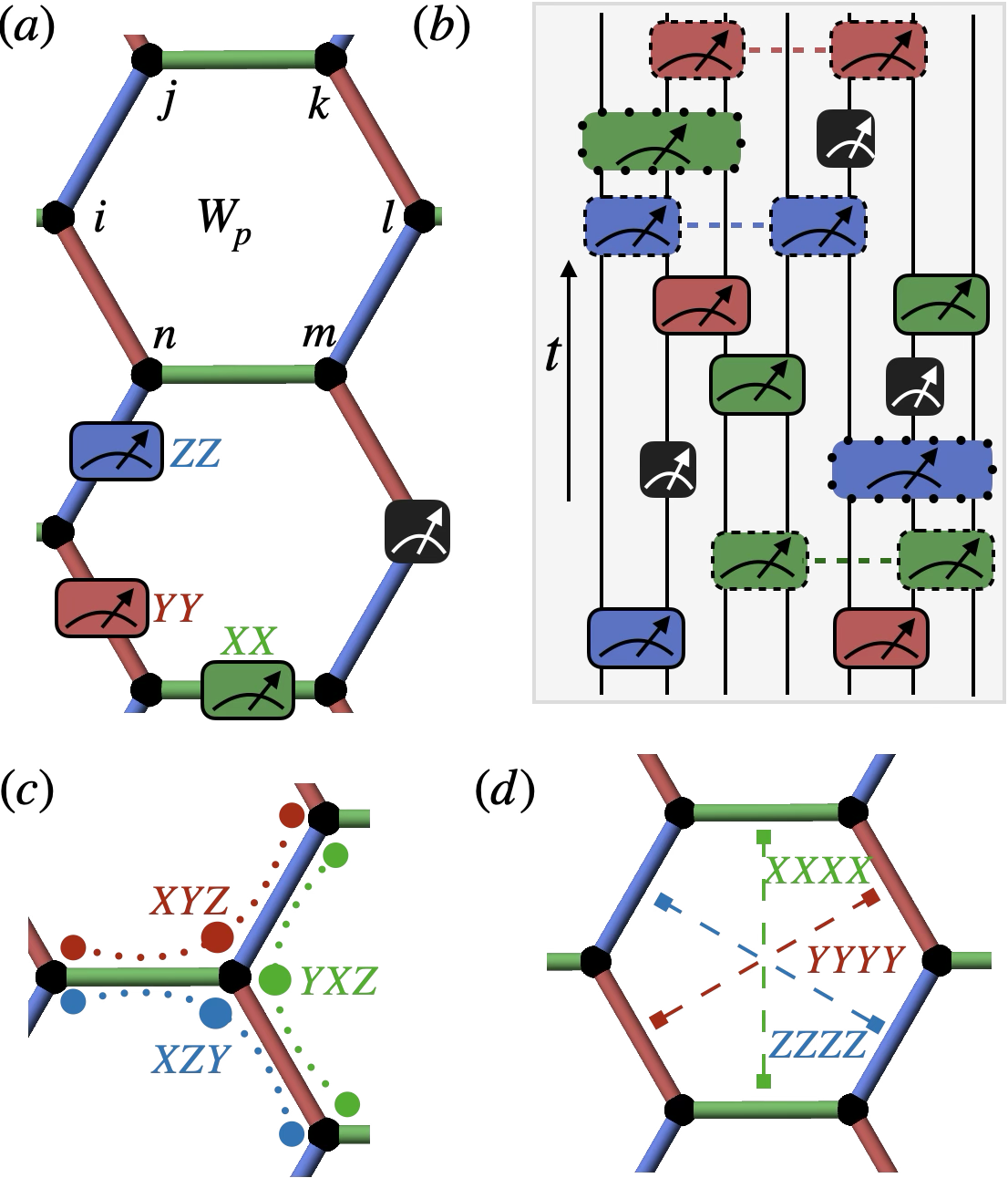}
    \caption{Schematic of our circuit model. (a) Two-qubit operators act along bonds (red, green, blue) with single-qubit operator (black). The plaquette operator is marked by $W_p$ with its corresponding sites $i-n$. (b) Circuit schematic: at each timestep operators are chosen at random to be measured. Depending on the model, these can have weights one through four. (c) At each site three three-qubit operators can be defined (d) Four-qubit operators are created by a pair of same flavor opposite bonds within a plaquette.}
    \label{fig:1}
\end{figure}

Particularly, topology can be introduced in measurement-only circuits by considering commuting projector models that exhibit topological order, like the toric code Hamiltonian\cite{Kitaev_AnnPhys2006,Lavasani_PRL2021}. A natural extension involves considering models that show topological order away from the commuting projector limit. In this context, the Kitaev honeycomb model serves as a paradigmatic exactly solvable system exhibiting a quantum spin liquid (QSL) ground state\cite{Kitaev_AnnPhys2006}. Depending on the bond anisotropies, it shows a gapped $Z_2$ spin liquid phase and a gapless phase with Dirac cones. The model hosts fractionalized anyons and long-range entanglement, exhibiting topological order. While it is exactly solvable, its topological order is sensitive to perturbations. A standard example is a time-reversal symmetry breaking magnetic field. At weak strengths, such a field opens a gap in the Dirac spectrum, generating a non-Abelian chiral spin liquid. However, strong fields polarize the spins into a topologically trivial product state.

The monitored-circuit formulation of the Kitaev model leverages this operator structure by interpreting bond and plaquette terms as measurement operators in a subsystem code, where the operators don't commute with each other \cite{Bacon_PRA2006,Aliferis_PRL2007}. Deterministic, time periodic (Floquet) schedules exhibit dynamically generated logical qubits in area-law regimes \cite{Hastings_Quant2021,Gidney_Quantum2021faulttolerant,Haah_Quantum2022boundarieshoneycomb,Aasen_PRB2022,Paetznick_PRXQ2023,Zhu_arxiv2023}, while fully stochastic \cite{Lavasani_PRB2023,Sriram_PRB2023,Zhu_PRR2024,Vijayvargia_arxiv2025} protocols of random bond measurements realize a richer phase diagram with critical-law entangled Majorana liquid and topological area-law phases depending on the anisotropy of the probabilities. This highlights that the choice of measured operators, rather than temporal order, organizes the phase diagram.

With this motivation, we study three physically motivated parity check operators within a stochastic measurement-only protocol built from Kitaev bond operators. (i) Single-qubit (magnetic field) operators that do not commute with the flux operator drive the system away from the Majorana liquid entanglement phases: they produce a volume-law phase at intermediate measurement probabilities and a trivial product state at larger probabilities. (ii) A time reversal symmetry breaking three-qubit operator, inspired by the chiral interaction that gaps the Hamiltonian model, stabilizes the critical-law regime against area-law behavior. (iii) A four-qubit operator that does commute with all plaquettes which is a simultaneous measurement of two opposite, same flavor, bonds on a hexagon generates a distinct volume-law phase that preserves topological sectors, i.e. extensive entanglement coexists with conserved flux memory. Across these cases, operator algebra, especially (non-)commutation with the fluxes emerges as the organizing principle separating the various short and long range entanglement phases.

The remainder of the paper is organized as follows. Section II discusses the Kitaev model followed by specifying the single-, three-, and four-qubit operators, and outlines the stochastic circuit protocol and diagnostics (bipartite entanglement, tripartite mutual information, and dynamical purification).
Section III  presents entanglement phase diagrams for the additional measurement operators added to the measurement only scheme, emphasizing the stabilization of the critical regime by the three-qubit operators and the coexistence of volume-law entanglement with conserved flux sectors under commuting four-qubit operators. We conclude with a summary of our results and an outlook.

\section{Model}
We briefly review the Kitaev honeycomb model before discussing its monitored circuit implementation \cite{Kitaev_AnnPhys2006}. The model describes spin-1/2 degrees of freedom on a honeycomb lattice with bond-directional couplings,
\begin{equation}\label{eq:hamiltonian}
    H \;=\; -K_x \!\!\sum_{\langle ij\rangle_x}\! \sigma_i^x \sigma_j^x
             -K_y \!\!\sum_{\langle ij\rangle_y}\! \sigma_i^y \sigma_j^y
             -K_z \!\!\sum_{\langle ij\rangle_z}\! \sigma_i^z \sigma_j^z ,
\end{equation}
where \(\langle ij\rangle_\alpha\) denotes nearest neighbors on \(\alpha=x,y,z\) bonds, as shown in Fig.~\ref{fig:1} (a). The key observation which
leads to the exact solvability of the Kitaev model is based on the plaquette operators $W_p=\sigma^x_i\sigma^y_j\sigma^z_k\sigma^x_l\sigma^y_m\sigma^z_n$
which commute with one another and with the Hamiltonian, \([W_p,W_{p'}]=0\) and \([W_p,H]=0\). The Hilbert space thus decomposes into sectors labeled by the eigenvalues \(W_p=\pm 1\).

A standard solution proceeds by representing each spin as four Majorana fermions, \(\sigma_i^\alpha = i\, b_i^\alpha c_i\). Defining bond variables \(u_{ij}^\alpha = i\, b_i^\alpha b_j^\alpha=\pm 1\), one obtains a free Majorana problem in a static \(\mathbb{Z}_2\) gauge background,
\begin{equation}\label{eq:majorana_ham}
    H \;=\; i \sum_{\langle ij\rangle_\alpha} K_\alpha\, u_{ij}^\alpha\, c_i c_j ,
\end{equation}
with the flux on a hexagon given by $W_p=\prod_{(ij)\in p} u_{ij}^{\alpha}$. According to Lieb’s theorem \cite{Lieb_PRL1994}, the zero-flux sector \(W_p=+1\) minimizes the energy, so one convenient gauge sets \(u_{ij}^\alpha=+1\) on all bonds. Diagonalizing Eq.~\eqref{eq:majorana_ham} then yields two regimes set by the couplings \(K_\alpha\): for any permutation with \(K_\alpha \ge K_\beta + K_\gamma\) the ground state is a gapped Abelian spin liquid that maps to the toric code in the anisotropic limit\cite{Kitaev_AnnPhys2003}, while otherwise the ground state is a gapless spin liquid with Dirac cones \cite{Kitaev_AnnPhys2006}.

A magnetic field term can be added to the model in the following way: 
\begin{equation}
    H_h=\sum_ih^x\sigma_i^x+h^y\sigma_i^y+h^z\sigma_i^z
\end{equation}
Note that this term breaks time reversal symmetry and as $[W_p,H_h]\neq 0$, the plaquette conservation is lost and the fluxes become dynamical \cite{Rau_PRL2014}. This is due to the fact that the itinerant $c$ Majorana fermions hybridize with gauge $b^\alpha$ Majorana fermions.
For large fields, $|h|\gg K^\alpha$, the ground state adiabatically connects to a trivial fully polarized phase which has no fractionalization or topological order. Depending on the orientation of the field, for weak field strengths, $(|h|\ll K^\alpha)$, the model either remains gapless or when $h^xh^yh^z\neq0$, at third order in perturbation, the effective Hamiltonian is given by
\begin{equation}
    H_3=-h_3\sum_{\langle ijk\rangle}\sigma^x_i\sigma^y_j\sigma^z_k + \text{cyclic permutations}
    \label{eq:p3}
\end{equation}
where $h_3\sim h^xh^yh^z/K^2$. Interestingly, this term commutes with the fluxes and hence preserves the fractionalization and topological order \cite{Kitaev_AnnPhys2006}. It is useful to represent this in the Majorana language: $H_3=h_3\sum_{\langle ijk\rangle}iu_{ij}u_{jk}c_ic_k$. This next-nearest neighbor Majorana hopping acts like a Haldane mass term and gaps the Dirac cones. The Majorana bands attain a Chern number $\nu=\pm 1$ which yields a non-Abelian chiral spin liquid.

In addition, we also consider a term that involves four qubits,
\begin{equation}
H_4 = h_4 \sum_{\alpha\in\{x,y,z\}} \sum_{\langle ij\rangle_\alpha}
\left(\sigma_i^\alpha \sigma_j^\alpha\right)\!
\left(\sigma_l^\alpha \sigma_m^\alpha\right),
\label{eq:p4}
\end{equation}
where $\langle ij\rangle_\alpha$ and $\langle lm\rangle_\alpha$ are opposite $\alpha$-bonds on the same hexagon [Fig.~\ref{fig:1}(d)].
 In terms of Majorana fermions this becomes
\(H_4=h_4\sum_{\alpha}\sum_{\langle ij\rangle_\alpha} u^\alpha_{ij}u^\alpha_{lm}\, c_i c_j c_l c_m\).
Unlike the operators considered earlier, \(H_4\) cannot be reduced to bilinears even after fixing the gauge. It also commutes with all plaquette fluxes, \([W_p,H_4]=0\), so flux sectors are preserved. In the zero-flux gauge \(u^\alpha_{ij}=+1\), \(H_4\) is a sum of quartic terms coupling opposite bonds on each hexagon. Since time reversal symmetry is unbroken, no Dirac mass is induced at leading order. We will refer to the resulting regime as an interacting Majorana liquid\cite{Zhu_PRR2024}. 

We will explore the effects of these three terms on the entanglement phase diagram of the monitored Kitaev model.

\subsection{Circuit dynamics}
We define the measurement protocol as follows: consider $N=2L^2$ qubits on the vertices of an $L\times L$ honeycomb lattice. Each edge of the lattice defines an operator, either $X_iX_j, Y_iY_j$, or $Z_iZ_j$, depending on the orientation of the edge, as defined above. We choose a bond direction $X,Y,Z$ according to probabilities $p_x,p_y,p_z$ (subject to $p_x+p_y+p_z=1$) then choose an operator of that type to measure uniformly at random. We define a time step to be $N$ such measurements. This is extended by considering the additional operators discussed above by adding them to the measurement protocol with probabilities $p_h/p_3/p_4$. As all terms consist of products of Pauli matrices this allows efficient simulation using the tableau formalism, for which we use the QuantumClifford.jl package \cite{Aaronson_PRA2004,Gottesman_arxiv1998}.

For $p_h$ measurements, we choose a site $i$ uniformly at random then choose either $X_i,Y_i,Z_i$ to measure at that site. As shown in Figure \ref{fig:1}, the three site operators can be labeled by their central site and the Pauli matrix at that site. Therefore for $p_3$ measurements, we sample them by choosing a central site, and then the three possible operators at that site. For $p_4$ we choose a bond (of any direction) at random, choose one of the two plaquettes it makes up, then create the four Pauli operator from the chosen bond and the same type bond across the chosen plaquette.

After a number of time steps polynomial in $L$, we determine the phase of our state using the tripartite mutual information. To define the tripartite mutual information we first define bipartite mutual information of regions $A$ and $B$, the information contained in the total region $AB$ that is not localizable to either region,
\begin{equation}
    I_2(A:B) = S_A + S_B - S_{AB}
\end{equation}
then the tripartite mutual information for regions $A,B,C$ is defined as
\begin{align}
    I_3(A:B:C) =& I_2(A:B) + I_2(B:C) - I_2(A:BC) \\
     =& S_A + S_B + S_C - S_{AB} - S_{BC} - S_{AC} \nonumber\\
     &+ S_{ABC}.
\end{align}
where $S_X$ is the entanglement entropy for region $X$. We partition the torus into four equal, non-contractible cylinders $A,B,C,D$ (Fig. \ref{fig:tpmi} inset) and evaluate $I_3(A\!:\!B\!:\!C)$ on three of them. This detects each different phase we are interested in as follows: the trivial entanglement phase shows no long range entanglement and therefore $I_3=0$. The topological area-law exhibits $I_3=1$, which is a contribution of one effective bell-pair shared between regions $A$ and $C$. This effective bell-pair can be constructed from the product of plaquettes and short strings. For critical-law, $I_3=-1$, which can be understood as the contribution from one independent long string shared between regions $ABC$ \cite{Sriram_PRB2023,Lavasani_PRB2023,Vijayvargia_arxiv2025v2}. The volume-law shows extensive entanglement, resulting in an extensive behavior in $I_3$ \cite{Zhu_PRR2024,Vijayvargia_arxiv2025v2}.

Although there is a finite width of crossover between these values due to finite size effects, the point of transition can be quantitatively determined using finite size scaling. There is a data collapse with a general form $I_3 = F[(r-r_c) L^{1/\nu}]$ for some transition radius $r_c$, corresponding to the crossing point of the curves, and scaling exponent $\nu$ similar to scaling observed in the monitored Kitaev model \cite{Lavasani_PRB2023,Klocke_PRB2025}. 

\begin{figure*}
    \centering
    \includegraphics[width=1\linewidth]{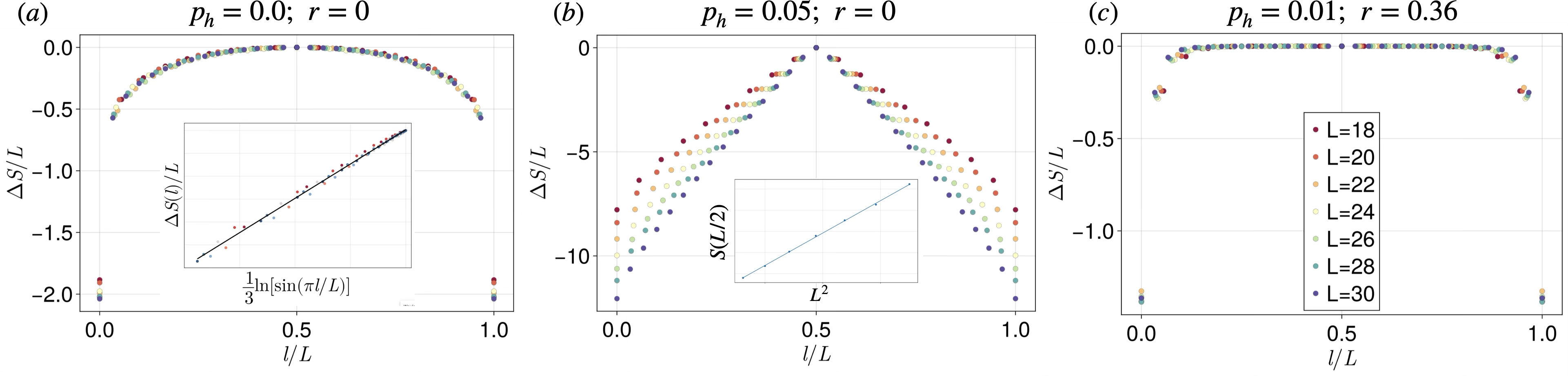}
    \caption{Scaling of the entanglement entropy with region size $\ell$ where $\Delta S$ measures the difference from the half-system entropy $\Delta S(\ell) = S(L/2)-S(\ell)$ for (a) critical-law, (b) volume-law, and (c) area-law phases.}
    \label{fig:phent}
\end{figure*}

We can compare this probe to other measures of an entanglement phase, such as the entanglement entropy scaling and the entanglement entropy of an $\ell\times L$ cylinder normalized to system size. This shows distinct shapes for the critical- volume- and area-law phases and, by fitting multiple functional forms, allows the determination of sub-leading contributions to overall entropy scaling. However, it is harder to determine the exact transition point between phases due to contributions from both phases near a phase boundary. We nevertheless verify the form of this scaling in the bulk of our phases in Fig \ref{fig:phent}.

Even though plaquettes are not measured directly, they can still be measured across multiple timesteps. The plaquettes can be defined as a product of bonds, as above. Each bond will be individually measured, then these can be combined by the update rules of the stabilizer simulation into a single plaquette operator. The rate of these effective measurements will play an important role when considering the phase diagram for the single-site operator.
After the plaquettes are measured, their values will be stable as they commute with all bond operators. Therefore, we will simplify our simulations by directly measuring all plaquettes as a first step, before continuing stochastic measurements as above. 

In addition to studying entanglement phase transitions in their own right, such systems can also define quantum error correcting codes when they protect logical qubits.
If we take the plaquettes as stabilizers defining a stabilizer code, where the measurement of the plaquettes provide a syndrome to correct certain errors, then this code contains two logical qubits when the system is taken to have periodic boundary conditions, corresponding to the two non-contractible loops. This can be brought closer to our bond measurement model by taking the two-qubit bond operators as checks making up a subsystem code. However, when taken as a subsystem code there are no logical qubits preserved \cite{Sriram_PRB2023}. This issue is resolved by introducing non-uniform time dynamics. A periodic Floquet schedule of cyclically measuring all $x$ then $y$ then $z$ bonds again preserves the two logical qubits from the non-contractible loops \cite{Hastings_Quant2021}. This generalizes to the stochastic setting we take here. Our model is able to measure the non-contractible loops, but only on timescales exponentially long in the system size. So the area-law phase here can be treated as a dynamic error correcting code preserving two logical qubits for exponentially long times. 

After the fluxes have been fixed by the dynamics, the remaining degrees of freedom are all in the Majorana fermion dynamics. These Majorana dynamics can be mapped to classical loop models by tracking the entanglement between pairs of Majoranas. As long as all terms are quadratic in the Majoranas, this mapping is exact \cite{Klocke_PRX2023}. These loop models can be treated with classical statistical mechanics, allowing both the classification of universality classes from system symmetries as well as much larger simulations permitting numerical verification of the critical exponents. See \cite{Klocke_PRB2025} for a detailed treatment. These methods show that, when limited to only quadratic hopping operators, the only accessible phases for these loop models are area-law or critical entanglement entropy scaling. The long time dynamics of these phases differ by the distribution of loop lengths. In the area-law phase there are only short range loops with long range ones exponentially suppressed. This leads to the characteristic area law entanglement growth, where only loops originating near the boundary of a region contribute to its entanglement entropy. The critical-law is characterized by an isotropic random walk of the loop endpoints leading to a power-law distribution of lengths below the system size. This heavier tailed distribution leads to the logarithmic correction to entanglement entropy scaling characteristic of the critical-law phase.
Achieving a volume-law phase, however, necessarily requires higher-order Majorana interactions. At small measurement rates these can potentially be treated perturbatively, but beyond a certain point they lead to the breakdown of the mapping to the loop model. 

We relate these ideas to our three additional operators. The single-qubit operator does not commute with the fluxes, so it will lead to a breakdown of the flux purification. Of course the $p_h\to 1$ limit is a trivial product state, but in the intermediate region we expect to see a volume-law phase. The three-qubit operator corresponds to next-nearest-neighbor hopping of the Majoranas, maintaining the loop model. Since this allows a faster diffusion of loop endpoints, we expect this to favor long loops and hence the critical-law phase. The four-qubit operator generates a four-Majorana interaction, also breaking the mapping to the loop model. Again generically we expect this to result in volume-law behavior. However, because this operator commutes with all the plaquette terms, this will only generate volume-law behavior in the itinerant Majorana degrees of freedom and not disturb the topological protection of the logical qubits. 

\section{Results}

\begin{figure}
    \centering
    \includegraphics[width=\linewidth]{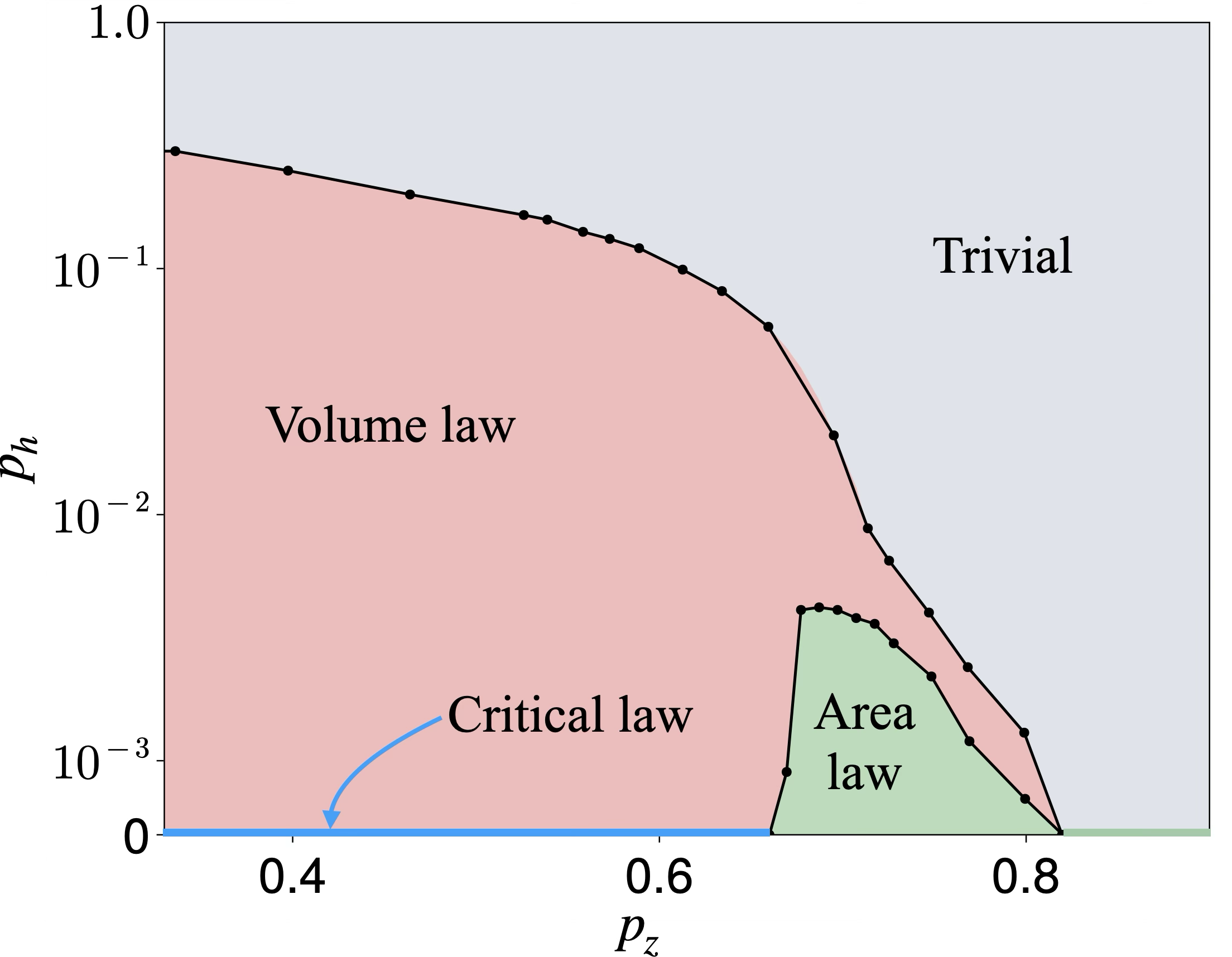}
    \caption{Phase diagram of $p_h,p_z$ (note logarithmic axis for $p_h$) showing area-law, volume-law, critical, and trivial phases.}
    \label{fig:ph}
\end{figure}

Next, we present phase diagrams for the monitored Kitaev model upon the addition of the single-, three-, and four-qubit terms described above. Phase boundaries are determined by the crossing point of the finite-size scaling of the tripartite mutual information, $I_3(A:B:C)$, as seen in Fig. \ref{fig:tpmi}.

Starting with the single-qubit term, Fig. \ref{fig:ph} shows the phase of the system under varying $p_z,p_h$, while setting $p_x=p_y=(1-p_z-p_h)/2$. For $p_h=0$, there is a critical phase for $p_z\leq0.66$ and an area-law phase above that. Starting from the critical-law phase, for any resolvable nonzero $p_h$, the steady state crosses over to a volume-law phase. Upon further increasing $p_h$ to values of order $p_h \sim 0.3$, the system transitions into the trivial product state.
The trivial state here corresponds to the magnetically polarized phase of the Kitaev Hamiltonian in the large-field limit. By contrast, the intervening volume-law phase has no counterpart in the ground state phase diagram of the equilibrium Kitaev model with a Zeeman field: in the Hamiltonian setting the system passes through a gapped non-Abelian chiral spin liquid window rather than a volume-law phase. Here the volume-law regime is a genuinely non-equilibrium feature of the monitored, stochastic dynamics, arising from repeated single-site measurements that do not commute with the plaquette operators and thus destroy flux conservation.

\begin{figure}
    \centering
    \includegraphics[width=\linewidth]{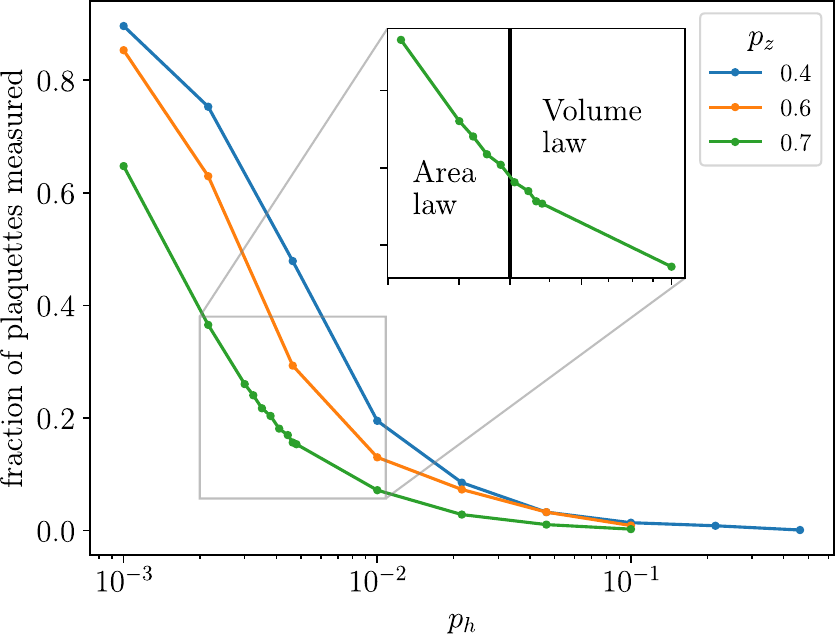}
    \caption{Fraction of plaquettes in span of stabilizer generators as a function of $p_h$ for selected $p_z$. Inset shows data near area/volume-law transition for $p_z=0.7$, with transition point marked by vertical line.}
    \label{fig:plaquette_fraction}
\end{figure}

Physically, there is a competition between bond measurements, which tend to project onto eigenstates of the plaquette operators, and single-site measurements, which randomize the neighboring fluxes and prevent plaquettes from remaining stabilizers of the late-time state.
 This can be controlled either by increasing $p_h$, which has a direct effect, or by increasing $p_z$, moving away from the isotropic $p_x=p_y=p_z$ point which decreases the effective rate of plaquette measurements. The direct effect of $p_h$ is illustrated in Fig. \ref{fig:plaquette_fraction}, which shows a decrease in the fraction of plaquettes spanned by the stabilizer generators, i.e. the fraction of plaquettes with measured expectation values, as $p_h$ increases, going to zero at the transition to the trivial phase. 

Above $p_z=0.82$, the measurement rate of the plaquettes is sufficiently small that the system goes to the trivial state for any value of $p_h$.

At $p_h=0$ and with $0.66 < p_z <0.82$, the system is in an area-law state. In contrast to the critical-law below $p_z=0.66$, there are no long stabilizer elements. Because of this at low $p_h$ the system doesn't immediately transition to a volume-law. The area-law region has a maximum value of $p_h=0.0042$ almost immediately above $p_z=0.68$. Even in this state there is a decrease in the fraction of plaquette operators fixed, as seen in Fig. \ref{fig:plaquette_fraction}. Indeed, there is no signature of the area-law/volume-law transition in the fraction of plaquettes which remain measured, as seen in the figure inset.

\begin{figure}
    \centering
    \includegraphics[width=\linewidth]{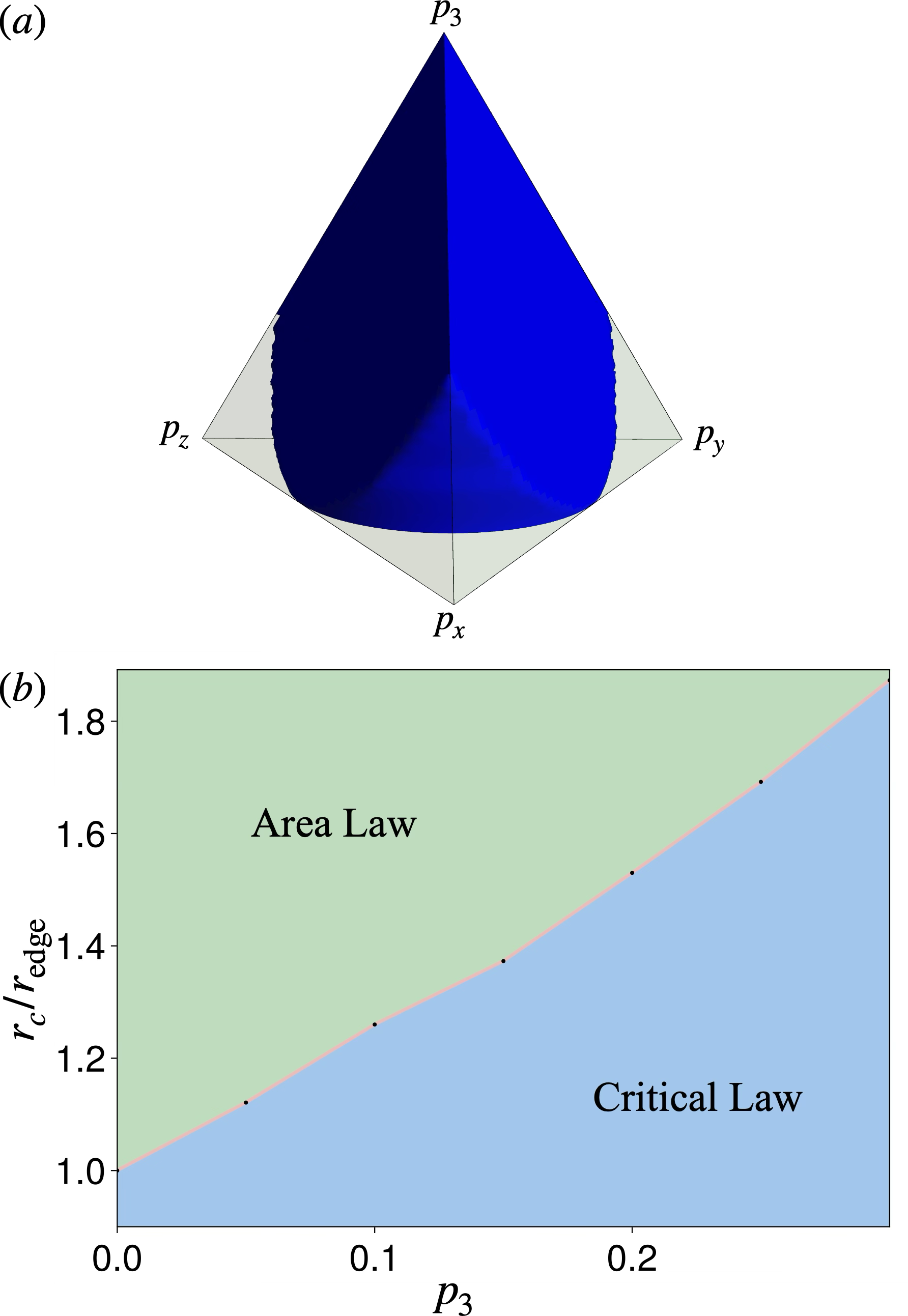}
    \caption{Phase diagram in the presence of the three-qubit operator. The critical-law region expands as a function of $p_3$. (a) Tetrahedron representing the full parameter space. (b) Normalized radius of the circular phase boundary plotted as a function of $p_3$.}
    \label{fig:p3}
\end{figure}
Next, we discuss the addition of operators that commute with the fluxes and do not destroy the topological order. As discussed above, a weak magnetic field perturbation leads to an effective Hamiltonian at third order of perturbation that has three-qubit operators: $\sigma^x_i\sigma^y_j\sigma^z_k$. These three-qubit operators can be thought of as the set of operators generated by multiplying two bond operators that share exactly one site, as shown in Fig. \ref{fig:1} (c). We include the measurement of these operators to the measurement-only scheme with a probability $p_3$. In terms of Majorana fermions, this term leads to a next-nearest neighbor hopping term. Therefore, the dynamics can still be mapped to a free Majorana loop model, for which only area-law and critical-law phases are expected \cite{Klocke_PRX2023,Zhu_PRR2024}.

\begin{figure}
    \centering
    \includegraphics[width=\linewidth]{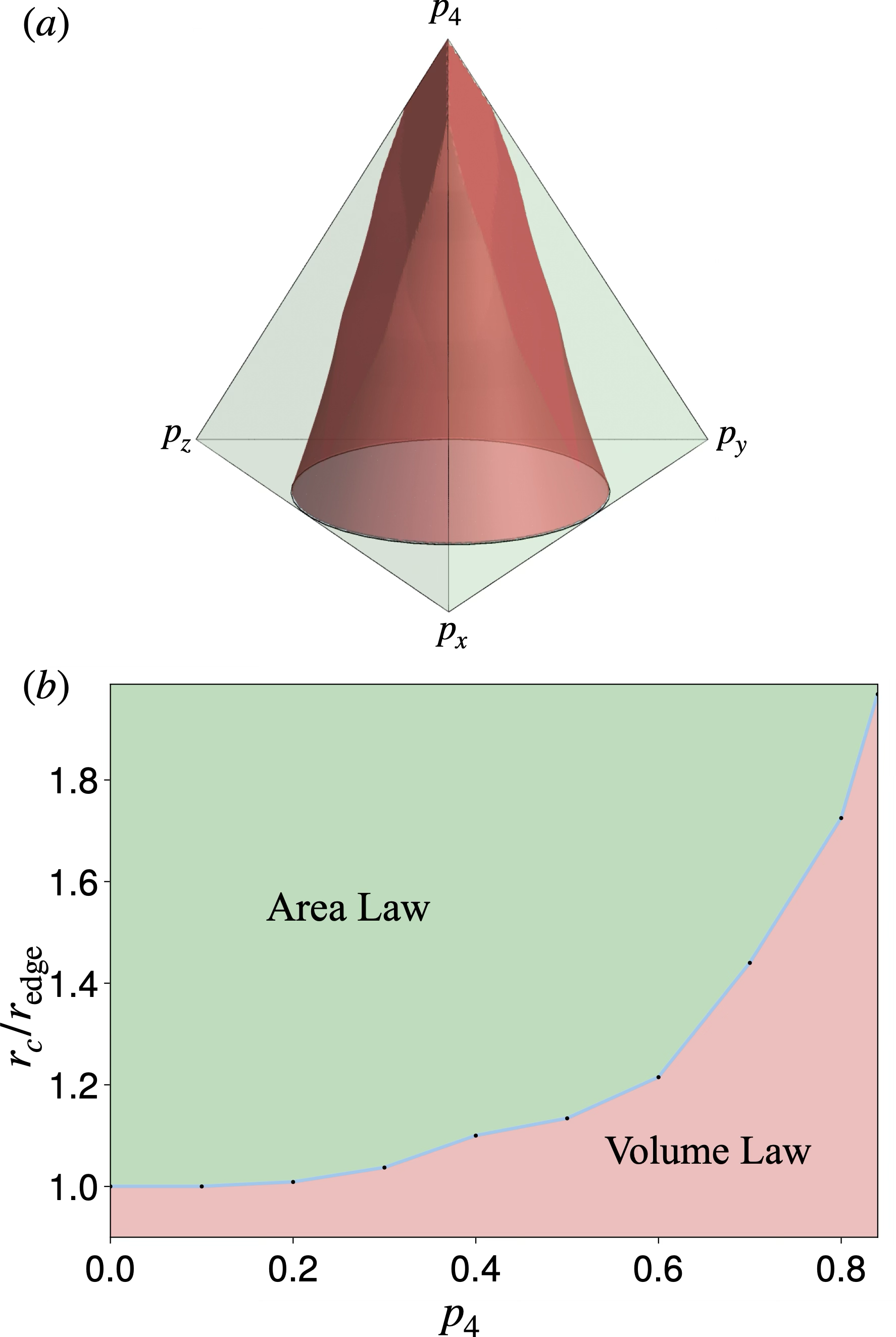}
    \caption{(a) Tetrahedral phase diagram of the four-qubit operator. The critical-law region in the $p_4=0$ plane turns to a volume-law region. (b) Normalized radius of the circular phase boundary plotted as a function of $p_4$.}
    \label{fig:p4}
\end{figure}
As the net result of measuring a three-qubit operator is to measure two different bond operators simultaneously, this leads to longer Pauli strings to be added to the stabilizer group than the case where only one type of disconnected bond is being measured. Resultantly, it is expected that for a higher $p_3$, the critical-law phase is favored over the area-law phase. This is what is observed in the phase diagram in Fig. \ref{fig:p3} (a). For constant $p_3$ slices of the tetrahedral phase diagram, we have found that the phase boundary is rotationally symmetric (see Appendix \ref{app:B}, Fig. \ref{fig:sphsym}). Therefore, in Fig. \ref{fig:p3} (b), we report the phase boundary as the radial distance from the isotropic point of the constant $p_3$ slices of the tetrahedron. The radius (within a constant $p_3$ slice) can be defined as $r = ||(p_x,p_y,p_z) - (\frac{1}{3},\frac{1}{3},\frac{1}{3})||$. The radius at the transition point, $r_c,$ is then normalized by the distance from the isotropic point to the edge-center $r_{\mathrm{edge}}$ such that for $p_3=0$ we have $r_c/r_{\mathrm{edge}}=1$. As seen in Fig. \ref{fig:p3} (b), the phase boundary, denoted by $r_c/r_{\mathrm{edge}}$ exhibits a monotonic trend of increasing expanse of the critical-law phase with increasing $p_3$.

It is instructive to contrast these circuit phases with the equilibrium phase diagram of the Kitaev Hamiltonian in a magnetic field. In the Hamiltonian problem, a uniform field with components along all three spin axes generates, at third order in perturbation theory, an effective chiral three-spin interaction that gaps the ground state and produces a chiral spin liquid with non-abelian anyonic excitations. In this intermediate field window the bulk spectrum is fully gapped, visons remain gapped, and the low energy theory is characterized by chiral Majorana edge modes and a threefold topological ground state degeneracy on the torus, before the system eventually polarizes into a trivial product state at large fields. In our measurement-only dynamics we do not find any analogue of this intermediate chiral phase. For the single-qubit measurements with probability $p_h$, the system crosses from the critical-law or area-law regimes directly into a non-topological volume-law phase and then into the trivial phase. This volume law originates from the fact that the single-site operators do not commute with the plaquette operators, so they constantly create and annihilate fluxes and add long Pauli strings to the stabilizers instead, rapidly destroying the plaquette based topological order. By contrast, the three-qubit measurements at probability $p_3$ preserve the plaquette fluxes and can be reformulated in terms of a free Majorana loop model, as discussed earlier. As a result, the entanglement phases do not realize non-Abelian topological order. In this sense, the monitored implementations of the field-induced operators reorganize the entanglement structure in ways that are qualitatively distinct from the chiral spin liquid window of the equilibrium Hamiltonian.

Subsequently, we consider a four-qubit operator composed of two bond operators, similar to the three-qubit case. However, here the bonds do not share a site and are of the same flavor within a single plaquette as depicted in Eq.~\ref{eq:p4}. This generates a quartic Majorana interaction
$u_{ij}^{\alpha}u_{lm}^{\alpha}c_i c_j c_l c_m$ without disturbing the plaquette stabilizer sector. As the measurement probability $p_4$ is increased from the base plane limit ($p_4=0$), the entanglement crosses over from the free-fermion critical form to a stable volume-law regime, consistent with an interacting Majorana liquid picture \cite{Zhu_PRR2024}. In contrast to six-Majorana plaquette measurements that can drive the system into a color code phase at large interaction strength, here we find no evidence for emergent color code order: the four-qubit measurements lead to an extensive number of long Pauli strings, which enhances information scrambling and hence the entanglement, but does not stabilize additional topological qubits.  Quantitatively, tripartite mutual information grows extensively, indicating an extensive set of conserved modes without gaps analogous to an interacting Fermi liquid in the monitored-circuit setting \cite{Zhu_PRR2024}. We construct a tetrahedral phase diagram in a way similar to that in the three-qubit case above. 
The tetrahedral phase diagram consists of a volume-law region controlled by $p_4$, and the radial cut $r_c/r_{\mathrm{edge}}$ increases monotonically with $p_4$, marking the continuous growth of the interacting-liquid domain as shown in Fig.~\ref{fig:p4}. 

Note that the volume-law observed in this phase is distinct from the volume-law obtained by single-site measurements with probability $p_h$. This can be understood by noting that in Fig. \ref{fig:plaquette_fraction}, the fraction of plaquette stabilizers in the steady state decreases monotonically as a function of $p_h$. However, this fraction stays at $1$ for the volume-law obtained from the four qubit measurement operator, as this operator commutes with the fluxes. Importantly, it also does not affect the logical qubits stored in topologically non-trivial cycles of the torus and this volume-law phase can be used to protect logical information, which is not possible in the volume-law phase obtained by the single site measurements. 

\section{Conclusion}
We studied the entanglement phase diagrams of the monitored Kitaev model with additional terms, motivated by similar perturbative terms in the Kitaev Hamiltonian. A single-qubit operator, analogous to a magnetic field, immediately induces a volume law phase in the \(p_h=0\) critical region, while preserving the area law up to some finite $p_h$. Further increasing this magnetic field's strength leads to a trivial product state. Adding the three- and four-qubit operators that commute with the plaquettes leads to a reduction of the area-law phase. In the case of the three-qubit operator, which is quadratic in the Majorana fermions, the stability of the critical phase is enhanced compared to the area-law phase. The four-qubit term immediately transforms the critical-law region to volume law which then expands as $p_4$ increases. In both cases the topological order protected by the plaquettes is preserved, as the additional operator commutes with the plaquettes. Therefore, the fast-scrambling effects characteristic of volume-law phases only apply to the Majorana degrees of freedom and the plaquettes continue to encode two topologically protected logical qubits. Interesting future directions include exploring higher-spin generalizations of the Kitaev model and the effects of single-ion anisotropy.
 
\section{Acknowledgments}
AV and OE acknowledge support from NSF Award No. DMR-2234352. EDR acknowledges support from NSF Award No. DMR-2206987. We thank the ASU Research Computing Center for high-performance computing resources.

TK, AV and EDR contributed equally to this work.

\appendix
\section{Entanglement scaling and tripartite mutual information}
The scaling of bipartite entanglement depends sensitively on the phase. 
Fig.~\ref{fig:phent}(a\textendash c) displays the entanglement entropy for cylindrical bipartitions of size $\ell \times L$, normalized to the $L/2 \times L$ cut, as a function of $\ell$. 
In the area\textendash law regime, long strings are exponentially rare, which confines noticeable deviations from the half\textendash system value to a narrow boundary layer of width $\mathcal{O}(\ln L)$. 
At criticality, the loop\textendash length distribution becomes algebraic, producing an overall growth $S \sim L \ln L$ and extending the range of $\ell$ over which the entropy differs from $S(L/2)$. 
In this regime the data collapse is well captured by the Calabrese\textendash Cardy form
\begin{equation}
  \frac{S}{L} \;=\; a(L) \,+\, b(L)\, L \,\ln\!\left[\sin\!\left(\frac{\pi \ell}{L}\right)\right],
\end{equation}
as illustrated in the inset of Fig.~\ref{fig:phent}(a).

The volume\textendash law phase exhibits an approximately linear dependence of $S(\ell)$ for intermediate cuts $\ell \!\sim\! L/4$, together with a cusp near $\ell \!=\! L/2$ that matches the Page curve expectation~\cite{Page1993}. 
Data from different $L$ do not collapse in this phase, and the inset highlights the $L^2$ scaling of the half system entropy, consistent with extensive entanglement.

While the entropy curves enable quantitative estimates of subleading contributions by fitting mixtures of the functional forms above, this same flexibility complicates precise boundary identification. 
Close to phase transitions, distinguishing a very small fitted component from an absent one is difficult, which limits the reliability of using entropy fits alone to pin down exact phase boundaries.

\noindent
These considerations motivate using the tripartite mutual information as the primary locator of phase boundaries. In the four-cylinder geometry already defined, $I_3$ varies smoothly across phases, taking values near $-1$ in the critical regime, approaching $+1$ in the area-law regime, and becoming extensive in the volume-law regime where its magnitude grows with $L$ (see Fig.~\ref{fig:phent}). Because boundary-law and short-range pieces cancel by construction, $I_3$ exhibits size-independent crossings that directly pin down the transition without delicate multi-parameter entropy fits. So curves for different $L$ cross at $r=r_c$ and collapse under the scaling variable, consistent with monitored Kitaev studies~\cite{Lavasani_PRB2023,Klocke_PRB2025,Vijayvargia_arxiv2025v2}. In practice this yields cleaner and more stable estimates of $r_c$ than entanglement-scaling analyses, while entropy-based diagnostics remain useful as secondary checks.

\begin{figure}
    \centering
    \includegraphics[width=0.7\linewidth]{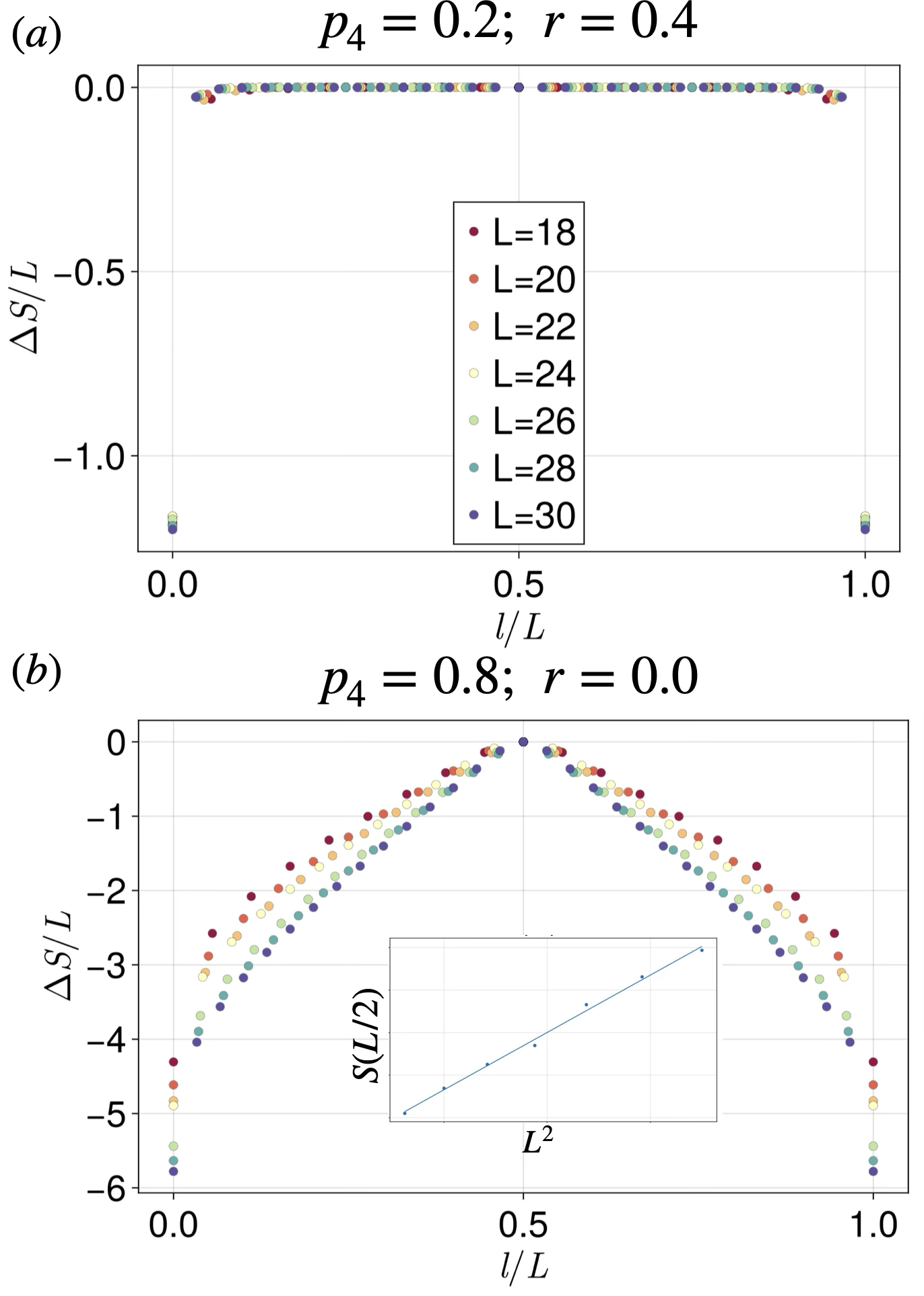}
    \caption{Scaling of entanglement entropy for finite values of $p_4$ for (a) area-law and (b) volume-law phases.}
    \label{fig:p4ent}
\end{figure}
\begin{figure}
    \centering
    \includegraphics[width=0.9\linewidth]{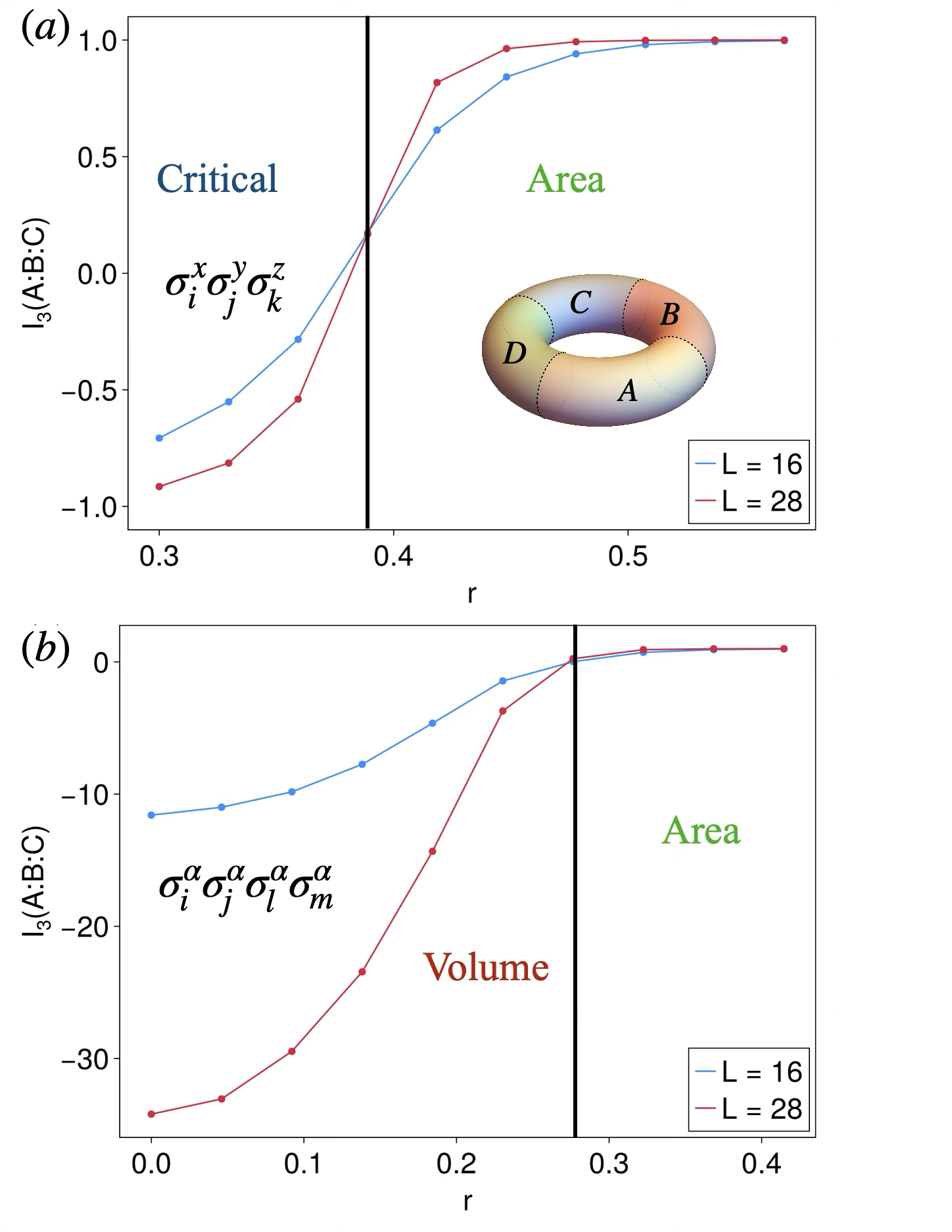}
    \caption{Quantitative determination of the phase boundary is done by finite size scaling of tri-partite mutual information. Curves of different system size $L$ will all intersect at the $L\to\infty$ limit. This is seen for both the three-qubit term (a) and four-qubit term (b). Inset (a): partitioning of the qubit lattice in four non-contractible sections to calculate tri-partite mutual information.}
    \label{fig:tpmi}
\end{figure}

\section{Rotational symmetry of the phase diagrams}
\label{app:B}
\begin{figure}
    \centering
    \includegraphics[width=1\linewidth]{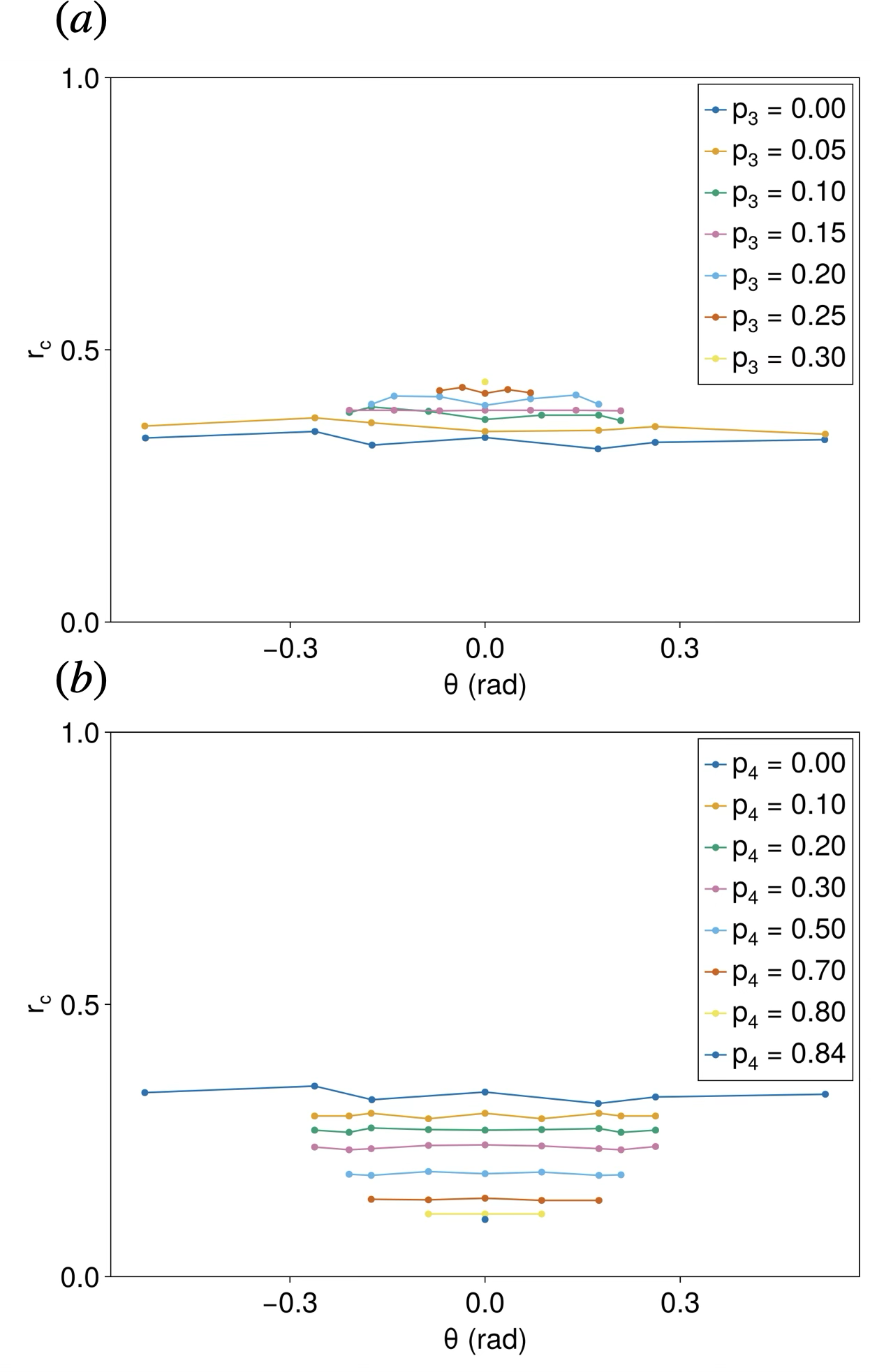}
    \caption{For all values of $p_3$ and $p_4$ the phase boundary occurs at a constant radius from the isotropic point.}
    \label{fig:sphsym}
\end{figure}

In the case of the three-qubit operator, taking a slice at constant $p_3,$ we follow a straightforward parametrization to calculate probabilities $p_i$ as
\begin{equation}
    p_i=\frac{1-p_3}{3}+ru_i(\theta), \quad i\in\{x,y,z\} 
\end{equation}
where $u_i$ are just directional weights such that $\sum_iu_i=0.$ By choosing a direction for $\theta$, it is simple to set $u_i(\theta)=\cos(\theta), \ \cos(\theta\pm2\pi/3).$ In the implementation, the distance from the isotropic point is bounded $(r\leq r_\text{max})$ to ensure that $p_i\in[0,1).$ This bound is calculated as $r_\text{max}(\theta,p_3)=\text{min}_{i:,u_i<0}\frac{(1-p_3)/3}{|u_i(\theta)|},$ the first contact at any edge where a $p_i$ vanishes. The same idea follows for the four-qubit term. \\
To locate the phase boundaries, we retrieve our $(p_x, p_y,p_z)$ from the parametrization, evolve the monitored circuit, and calculate the tripartite mutual information, averaged over 100 runs.\\ 
Fig. \ref{fig:sphsym} shows that the transition point is independent of the direction $\theta,$ within numerical precision. This is true for all values of $p_3/p_4$. Note that near the edges or faces of the tetrahedron where one or more probability goes to zero it requires increasingly long time for the system to reach a steady state. Therefore we avoided such points, leading to the reduction in sampled angles for increased $p_3/p_4$, as the phase boundary moves nearer the edges of the tetrahedron. 

\bibliography{references.bib}

\begin{thebibliography}{35}%
\makeatletter
\providecommand \@ifxundefined [1]{%
 \@ifx{#1\undefined}
}%
\providecommand \@ifnum [1]{%
 \ifnum #1\expandafter \@firstoftwo
 \else \expandafter \@secondoftwo
 \fi
}%
\providecommand \@ifx [1]{%
 \ifx #1\expandafter \@firstoftwo
 \else \expandafter \@secondoftwo
 \fi
}%
\providecommand \natexlab [1]{#1}%
\providecommand \enquote  [1]{``#1''}%
\providecommand \bibnamefont  [1]{#1}%
\providecommand \bibfnamefont [1]{#1}%
\providecommand \citenamefont [1]{#1}%
\providecommand \href@noop [0]{\@secondoftwo}%
\providecommand \href [0]{\begingroup \@sanitize@url \@href}%
\providecommand \@href[1]{\@@startlink{#1}\@@href}%
\providecommand \@@href[1]{\endgroup#1\@@endlink}%
\providecommand \@sanitize@url [0]{\catcode `\\12\catcode `\$12\catcode `\&12\catcode `\#12\catcode `\^12\catcode `\_12\catcode `\%12\relax}%
\providecommand \@@startlink[1]{}%
\providecommand \@@endlink[0]{}%
\providecommand \url  [0]{\begingroup\@sanitize@url \@url }%
\providecommand \@url [1]{\endgroup\@href {#1}{\urlprefix }}%
\providecommand \urlprefix  [0]{URL }%
\providecommand \Eprint [0]{\href }%
\providecommand \doibase [0]{https://doi.org/}%
\providecommand \selectlanguage [0]{\@gobble}%
\providecommand \bibinfo  [0]{\@secondoftwo}%
\providecommand \bibfield  [0]{\@secondoftwo}%
\providecommand \translation [1]{[#1]}%
\providecommand \BibitemOpen [0]{}%
\providecommand \bibitemStop [0]{}%
\providecommand \bibitemNoStop [0]{.\EOS\space}%
\providecommand \EOS [0]{\spacefactor3000\relax}%
\providecommand \BibitemShut  [1]{\csname bibitem#1\endcsname}%
\let\auto@bib@innerbib\@empty
\bibitem [{\citenamefont {Skinner}\ \emph {et~al.}(2019)\citenamefont {Skinner}, \citenamefont {Ruhman},\ and\ \citenamefont {Nahum}}]{Skinner_PRX2019}%
  \BibitemOpen
  \bibfield  {author} {\bibinfo {author} {\bibfnamefont {B.}~\bibnamefont {Skinner}}, \bibinfo {author} {\bibfnamefont {J.}~\bibnamefont {Ruhman}},\ and\ \bibinfo {author} {\bibfnamefont {A.}~\bibnamefont {Nahum}},\ }\bibfield  {title} {\bibinfo {title} {Measurement-induced phase transitions in the dynamics of entanglement},\ }\href {https://doi.org/10.1103/PhysRevX.9.031009} {\bibfield  {journal} {\bibinfo  {journal} {Phys. Rev. X}\ }\textbf {\bibinfo {volume} {9}},\ \bibinfo {pages} {031009} (\bibinfo {year} {2019})}\BibitemShut {NoStop}%
\bibitem [{\citenamefont {Gullans}\ and\ \citenamefont {Huse}(2020)}]{Gullans_PRX2020}%
  \BibitemOpen
  \bibfield  {author} {\bibinfo {author} {\bibfnamefont {M.~J.}\ \bibnamefont {Gullans}}\ and\ \bibinfo {author} {\bibfnamefont {D.~A.}\ \bibnamefont {Huse}},\ }\bibfield  {title} {\bibinfo {title} {Dynamical purification phase transition induced by quantum measurements},\ }\href {https://doi.org/10.1103/PhysRevX.10.041020} {\bibfield  {journal} {\bibinfo  {journal} {Phys. Rev. X}\ }\textbf {\bibinfo {volume} {10}},\ \bibinfo {pages} {041020} (\bibinfo {year} {2020})}\BibitemShut {NoStop}%
\bibitem [{\citenamefont {Choi}\ \emph {et~al.}(2020)\citenamefont {Choi}, \citenamefont {Bao}, \citenamefont {Qi},\ and\ \citenamefont {Altman}}]{Choi_PRL2020}%
  \BibitemOpen
  \bibfield  {author} {\bibinfo {author} {\bibfnamefont {S.}~\bibnamefont {Choi}}, \bibinfo {author} {\bibfnamefont {Y.}~\bibnamefont {Bao}}, \bibinfo {author} {\bibfnamefont {X.-L.}\ \bibnamefont {Qi}},\ and\ \bibinfo {author} {\bibfnamefont {E.}~\bibnamefont {Altman}},\ }\bibfield  {title} {\bibinfo {title} {Quantum error correction in scrambling dynamics and measurement-induced phase transition},\ }\href {https://doi.org/10.1103/PhysRevLett.125.030505} {\bibfield  {journal} {\bibinfo  {journal} {Phys. Rev. Lett.}\ }\textbf {\bibinfo {volume} {125}},\ \bibinfo {pages} {030505} (\bibinfo {year} {2020})}\BibitemShut {NoStop}%
\bibitem [{\citenamefont {{Fisher}}\ \emph {et~al.}(2023)\citenamefont {{Fisher}}, \citenamefont {{Khemani}}, \citenamefont {{Nahum}},\ and\ \citenamefont {{Vijay}}}]{Fisher_annrev2023}%
  \BibitemOpen
  \bibfield  {author} {\bibinfo {author} {\bibfnamefont {M.~P.~A.}\ \bibnamefont {{Fisher}}}, \bibinfo {author} {\bibfnamefont {V.}~\bibnamefont {{Khemani}}}, \bibinfo {author} {\bibfnamefont {A.}~\bibnamefont {{Nahum}}},\ and\ \bibinfo {author} {\bibfnamefont {S.}~\bibnamefont {{Vijay}}},\ }\bibfield  {title} {\bibinfo {title} {{Random Quantum Circuits}},\ }\href {https://doi.org/10.1146/annurev-conmatphys-031720-030658} {\bibfield  {journal} {\bibinfo  {journal} {Annual Review of Condensed Matter Physics}\ }\textbf {\bibinfo {volume} {14}},\ \bibinfo {pages} {335} (\bibinfo {year} {2023})},\ \Eprint {https://arxiv.org/abs/2207.14280} {arXiv:2207.14280 [quant-ph]} \BibitemShut {NoStop}%
\bibitem [{\citenamefont {Zabalo}\ \emph {et~al.}(2022)\citenamefont {Zabalo}, \citenamefont {Gullans}, \citenamefont {Wilson}, \citenamefont {Vasseur}, \citenamefont {Ludwig}, \citenamefont {Gopalakrishnan}, \citenamefont {Huse},\ and\ \citenamefont {Pixley}}]{Zabalo_PRL2022}%
  \BibitemOpen
  \bibfield  {author} {\bibinfo {author} {\bibfnamefont {A.}~\bibnamefont {Zabalo}}, \bibinfo {author} {\bibfnamefont {M.~J.}\ \bibnamefont {Gullans}}, \bibinfo {author} {\bibfnamefont {J.~H.}\ \bibnamefont {Wilson}}, \bibinfo {author} {\bibfnamefont {R.}~\bibnamefont {Vasseur}}, \bibinfo {author} {\bibfnamefont {A.~W.~W.}\ \bibnamefont {Ludwig}}, \bibinfo {author} {\bibfnamefont {S.}~\bibnamefont {Gopalakrishnan}}, \bibinfo {author} {\bibfnamefont {D.~A.}\ \bibnamefont {Huse}},\ and\ \bibinfo {author} {\bibfnamefont {J.~H.}\ \bibnamefont {Pixley}},\ }\bibfield  {title} {\bibinfo {title} {Operator scaling dimensions and multifractality at measurement-induced transitions},\ }\href {https://doi.org/10.1103/PhysRevLett.128.050602} {\bibfield  {journal} {\bibinfo  {journal} {Phys. Rev. Lett.}\ }\textbf {\bibinfo {volume} {128}},\ \bibinfo {pages} {050602} (\bibinfo {year} {2022})}\BibitemShut {NoStop}%
\bibitem [{\citenamefont {{Lavasani}}\ \emph {et~al.}(2021)\citenamefont {{Lavasani}}, \citenamefont {{Alavirad}},\ and\ \citenamefont {{Barkeshli}}}]{Lavasani_NatPhys2021}%
  \BibitemOpen
  \bibfield  {author} {\bibinfo {author} {\bibfnamefont {A.}~\bibnamefont {{Lavasani}}}, \bibinfo {author} {\bibfnamefont {Y.}~\bibnamefont {{Alavirad}}},\ and\ \bibinfo {author} {\bibfnamefont {M.}~\bibnamefont {{Barkeshli}}},\ }\bibfield  {title} {\bibinfo {title} {{Measurement-induced topological entanglement transitions in symmetric random quantum circuits}},\ }\href {https://doi.org/10.1038/s41567-020-01112-z} {\bibfield  {journal} {\bibinfo  {journal} {Nature Physics}\ }\textbf {\bibinfo {volume} {17}},\ \bibinfo {pages} {342} (\bibinfo {year} {2021})},\ \Eprint {https://arxiv.org/abs/2004.07243} {arXiv:2004.07243 [quant-ph]} \BibitemShut {NoStop}%
\bibitem [{\citenamefont {Sang}\ and\ \citenamefont {Hsieh}(2021)}]{Sang_PRR2021}%
  \BibitemOpen
  \bibfield  {author} {\bibinfo {author} {\bibfnamefont {S.}~\bibnamefont {Sang}}\ and\ \bibinfo {author} {\bibfnamefont {T.~H.}\ \bibnamefont {Hsieh}},\ }\bibfield  {title} {\bibinfo {title} {Measurement-protected quantum phases},\ }\href {https://doi.org/10.1103/PhysRevResearch.3.023200} {\bibfield  {journal} {\bibinfo  {journal} {Phys. Rev. Res.}\ }\textbf {\bibinfo {volume} {3}},\ \bibinfo {pages} {023200} (\bibinfo {year} {2021})}\BibitemShut {NoStop}%
\bibitem [{\citenamefont {Klocke}\ and\ \citenamefont {Buchhold}(2022)}]{Klocke_PRB2022}%
  \BibitemOpen
  \bibfield  {author} {\bibinfo {author} {\bibfnamefont {K.}~\bibnamefont {Klocke}}\ and\ \bibinfo {author} {\bibfnamefont {M.}~\bibnamefont {Buchhold}},\ }\bibfield  {title} {\bibinfo {title} {Topological order and entanglement dynamics in the measurement-only {XZZX} quantum code},\ }\href {https://doi.org/10.1103/PhysRevB.106.104307} {\bibfield  {journal} {\bibinfo  {journal} {Phys. Rev. B}\ }\textbf {\bibinfo {volume} {106}},\ \bibinfo {pages} {104307} (\bibinfo {year} {2022})}\BibitemShut {NoStop}%
\bibitem [{\citenamefont {Ippoliti}\ \emph {et~al.}(2021)\citenamefont {Ippoliti}, \citenamefont {Gullans}, \citenamefont {Gopalakrishnan}, \citenamefont {Huse},\ and\ \citenamefont {Khemani}}]{Ippoliti_PRX2021}%
  \BibitemOpen
  \bibfield  {author} {\bibinfo {author} {\bibfnamefont {M.}~\bibnamefont {Ippoliti}}, \bibinfo {author} {\bibfnamefont {M.~J.}\ \bibnamefont {Gullans}}, \bibinfo {author} {\bibfnamefont {S.}~\bibnamefont {Gopalakrishnan}}, \bibinfo {author} {\bibfnamefont {D.~A.}\ \bibnamefont {Huse}},\ and\ \bibinfo {author} {\bibfnamefont {V.}~\bibnamefont {Khemani}},\ }\bibfield  {title} {\bibinfo {title} {Entanglement phase transitions in measurement-only dynamics},\ }\href {https://doi.org/10.1103/PhysRevX.11.011030} {\bibfield  {journal} {\bibinfo  {journal} {Phys. Rev. X}\ }\textbf {\bibinfo {volume} {11}},\ \bibinfo {pages} {011030} (\bibinfo {year} {2021})}\BibitemShut {NoStop}%
\bibitem [{\citenamefont {Lavasani}\ \emph {et~al.}(2021)\citenamefont {Lavasani}, \citenamefont {Alavirad},\ and\ \citenamefont {Barkeshli}}]{Lavasani_PRL2021}%
  \BibitemOpen
  \bibfield  {author} {\bibinfo {author} {\bibfnamefont {A.}~\bibnamefont {Lavasani}}, \bibinfo {author} {\bibfnamefont {Y.}~\bibnamefont {Alavirad}},\ and\ \bibinfo {author} {\bibfnamefont {M.}~\bibnamefont {Barkeshli}},\ }\bibfield  {title} {\bibinfo {title} {Topological order and criticality in $(2+1)\mathrm{D}$ monitored random quantum circuits},\ }\href {https://doi.org/10.1103/PhysRevLett.127.235701} {\bibfield  {journal} {\bibinfo  {journal} {Phys. Rev. Lett.}\ }\textbf {\bibinfo {volume} {127}},\ \bibinfo {pages} {235701} (\bibinfo {year} {2021})}\BibitemShut {NoStop}%
\bibitem [{\citenamefont {Hastings}\ and\ \citenamefont {Haah}(2021)}]{Hastings_Quant2021}%
  \BibitemOpen
  \bibfield  {author} {\bibinfo {author} {\bibfnamefont {M.~B.}\ \bibnamefont {Hastings}}\ and\ \bibinfo {author} {\bibfnamefont {J.}~\bibnamefont {Haah}},\ }\bibfield  {title} {\bibinfo {title} {Dynamically {G}enerated {L}ogical {Q}ubits},\ }\href {https://doi.org/10.22331/q-2021-10-19-564} {\bibfield  {journal} {\bibinfo  {journal} {{Quantum}}\ }\textbf {\bibinfo {volume} {5}},\ \bibinfo {pages} {564} (\bibinfo {year} {2021})}\BibitemShut {NoStop}%
\bibitem [{\citenamefont {Vu}\ \emph {et~al.}(2024)\citenamefont {Vu}, \citenamefont {Lavasani}, \citenamefont {Lee},\ and\ \citenamefont {Fisher}}]{Vu_PRL2024}%
  \BibitemOpen
  \bibfield  {author} {\bibinfo {author} {\bibfnamefont {D.}~\bibnamefont {Vu}}, \bibinfo {author} {\bibfnamefont {A.}~\bibnamefont {Lavasani}}, \bibinfo {author} {\bibfnamefont {J.~Y.}\ \bibnamefont {Lee}},\ and\ \bibinfo {author} {\bibfnamefont {M.~P.~A.}\ \bibnamefont {Fisher}},\ }\bibfield  {title} {\bibinfo {title} {Stable measurement-induced floquet enriched topological order},\ }\href {https://doi.org/10.1103/PhysRevLett.132.070401} {\bibfield  {journal} {\bibinfo  {journal} {Phys. Rev. Lett.}\ }\textbf {\bibinfo {volume} {132}},\ \bibinfo {pages} {070401} (\bibinfo {year} {2024})}\BibitemShut {NoStop}%
\bibitem [{\citenamefont {Agrawal}\ \emph {et~al.}(2022)\citenamefont {Agrawal}, \citenamefont {Zabalo}, \citenamefont {Chen}, \citenamefont {Wilson}, \citenamefont {Potter}, \citenamefont {Pixley}, \citenamefont {Gopalakrishnan},\ and\ \citenamefont {Vasseur}}]{Agrawal_PRX2022}%
  \BibitemOpen
  \bibfield  {author} {\bibinfo {author} {\bibfnamefont {U.}~\bibnamefont {Agrawal}}, \bibinfo {author} {\bibfnamefont {A.}~\bibnamefont {Zabalo}}, \bibinfo {author} {\bibfnamefont {K.}~\bibnamefont {Chen}}, \bibinfo {author} {\bibfnamefont {J.~H.}\ \bibnamefont {Wilson}}, \bibinfo {author} {\bibfnamefont {A.~C.}\ \bibnamefont {Potter}}, \bibinfo {author} {\bibfnamefont {J.~H.}\ \bibnamefont {Pixley}}, \bibinfo {author} {\bibfnamefont {S.}~\bibnamefont {Gopalakrishnan}},\ and\ \bibinfo {author} {\bibfnamefont {R.}~\bibnamefont {Vasseur}},\ }\bibfield  {title} {\bibinfo {title} {Entanglement and charge-sharpening transitions in {U(1)} symmetric monitored quantum circuits},\ }\href {https://doi.org/10.1103/PhysRevX.12.041002} {\bibfield  {journal} {\bibinfo  {journal} {Phys. Rev. X}\ }\textbf {\bibinfo {volume} {12}},\ \bibinfo {pages} {041002} (\bibinfo {year} {2022})}\BibitemShut {NoStop}%
\bibitem [{\citenamefont {Barratt}\ \emph {et~al.}(2022)\citenamefont {Barratt}, \citenamefont {Agrawal}, \citenamefont {Gopalakrishnan}, \citenamefont {Huse}, \citenamefont {Vasseur},\ and\ \citenamefont {Potter}}]{Barratt_PRL2022}%
  \BibitemOpen
  \bibfield  {author} {\bibinfo {author} {\bibfnamefont {F.}~\bibnamefont {Barratt}}, \bibinfo {author} {\bibfnamefont {U.}~\bibnamefont {Agrawal}}, \bibinfo {author} {\bibfnamefont {S.}~\bibnamefont {Gopalakrishnan}}, \bibinfo {author} {\bibfnamefont {D.~A.}\ \bibnamefont {Huse}}, \bibinfo {author} {\bibfnamefont {R.}~\bibnamefont {Vasseur}},\ and\ \bibinfo {author} {\bibfnamefont {A.~C.}\ \bibnamefont {Potter}},\ }\bibfield  {title} {\bibinfo {title} {Field theory of charge sharpening in symmetric monitored quantum circuits},\ }\href {https://doi.org/10.1103/PhysRevLett.129.120604} {\bibfield  {journal} {\bibinfo  {journal} {Phys. Rev. Lett.}\ }\textbf {\bibinfo {volume} {129}},\ \bibinfo {pages} {120604} (\bibinfo {year} {2022})}\BibitemShut {NoStop}%
\bibitem [{\citenamefont {Kitaev}(2006)}]{Kitaev_AnnPhys2006}%
  \BibitemOpen
  \bibfield  {author} {\bibinfo {author} {\bibfnamefont {A.}~\bibnamefont {Kitaev}},\ }\bibfield  {title} {\bibinfo {title} {Anyons in an exactly solved model and beyond},\ }\href {https://doi.org/https://doi.org/10.1016/j.aop.2005.10.005} {\bibfield  {journal} {\bibinfo  {journal} {Annals of Physics}\ }\textbf {\bibinfo {volume} {321}},\ \bibinfo {pages} {2} (\bibinfo {year} {2006})}\BibitemShut {NoStop}%
\bibitem [{\citenamefont {Bacon}(2006)}]{Bacon_PRA2006}%
  \BibitemOpen
  \bibfield  {author} {\bibinfo {author} {\bibfnamefont {D.}~\bibnamefont {Bacon}},\ }\bibfield  {title} {\bibinfo {title} {Operator quantum error-correcting subsystems for self-correcting quantum memories},\ }\href {https://doi.org/10.1103/PhysRevA.73.012340} {\bibfield  {journal} {\bibinfo  {journal} {Phys. Rev. A}\ }\textbf {\bibinfo {volume} {73}},\ \bibinfo {pages} {012340} (\bibinfo {year} {2006})}\BibitemShut {NoStop}%
\bibitem [{\citenamefont {Aliferis}\ and\ \citenamefont {Cross}(2007)}]{Aliferis_PRL2007}%
  \BibitemOpen
  \bibfield  {author} {\bibinfo {author} {\bibfnamefont {P.}~\bibnamefont {Aliferis}}\ and\ \bibinfo {author} {\bibfnamefont {A.~W.}\ \bibnamefont {Cross}},\ }\bibfield  {title} {\bibinfo {title} {Subsystem fault tolerance with the bacon-shor code},\ }\href {https://doi.org/10.1103/PhysRevLett.98.220502} {\bibfield  {journal} {\bibinfo  {journal} {Phys. Rev. Lett.}\ }\textbf {\bibinfo {volume} {98}},\ \bibinfo {pages} {220502} (\bibinfo {year} {2007})}\BibitemShut {NoStop}%
\bibitem [{\citenamefont {Gidney}\ \emph {et~al.}(2021)\citenamefont {Gidney}, \citenamefont {Newman}, \citenamefont {Fowler},\ and\ \citenamefont {Broughton}}]{Gidney_Quantum2021faulttolerant}%
  \BibitemOpen
  \bibfield  {author} {\bibinfo {author} {\bibfnamefont {C.}~\bibnamefont {Gidney}}, \bibinfo {author} {\bibfnamefont {M.}~\bibnamefont {Newman}}, \bibinfo {author} {\bibfnamefont {A.}~\bibnamefont {Fowler}},\ and\ \bibinfo {author} {\bibfnamefont {M.}~\bibnamefont {Broughton}},\ }\bibfield  {title} {\bibinfo {title} {A {F}ault-{T}olerant {H}oneycomb {M}emory},\ }\href {https://doi.org/10.22331/q-2021-12-20-605} {\bibfield  {journal} {\bibinfo  {journal} {{Quantum}}\ }\textbf {\bibinfo {volume} {5}},\ \bibinfo {pages} {605} (\bibinfo {year} {2021})}\BibitemShut {NoStop}%
\bibitem [{\citenamefont {Haah}\ and\ \citenamefont {Hastings}(2022)}]{Haah_Quantum2022boundarieshoneycomb}%
  \BibitemOpen
  \bibfield  {author} {\bibinfo {author} {\bibfnamefont {J.}~\bibnamefont {Haah}}\ and\ \bibinfo {author} {\bibfnamefont {M.~B.}\ \bibnamefont {Hastings}},\ }\bibfield  {title} {\bibinfo {title} {Boundaries for the {H}oneycomb {C}ode},\ }\href {https://doi.org/10.22331/q-2022-04-21-693} {\bibfield  {journal} {\bibinfo  {journal} {{Quantum}}\ }\textbf {\bibinfo {volume} {6}},\ \bibinfo {pages} {693} (\bibinfo {year} {2022})}\BibitemShut {NoStop}%
\bibitem [{\citenamefont {Aasen}\ \emph {et~al.}(2022)\citenamefont {Aasen}, \citenamefont {Wang},\ and\ \citenamefont {Hastings}}]{Aasen_PRB2022}%
  \BibitemOpen
  \bibfield  {author} {\bibinfo {author} {\bibfnamefont {D.}~\bibnamefont {Aasen}}, \bibinfo {author} {\bibfnamefont {Z.}~\bibnamefont {Wang}},\ and\ \bibinfo {author} {\bibfnamefont {M.~B.}\ \bibnamefont {Hastings}},\ }\bibfield  {title} {\bibinfo {title} {Adiabatic paths of hamiltonians, symmetries of topological order, and automorphism codes},\ }\href {https://doi.org/10.1103/PhysRevB.106.085122} {\bibfield  {journal} {\bibinfo  {journal} {Phys. Rev. B}\ }\textbf {\bibinfo {volume} {106}},\ \bibinfo {pages} {085122} (\bibinfo {year} {2022})}\BibitemShut {NoStop}%
\bibitem [{\citenamefont {Paetznick}\ \emph {et~al.}(2023)\citenamefont {Paetznick}, \citenamefont {Knapp}, \citenamefont {Delfosse}, \citenamefont {Bauer}, \citenamefont {Haah}, \citenamefont {Hastings},\ and\ \citenamefont {da~Silva}}]{Paetznick_PRXQ2023}%
  \BibitemOpen
  \bibfield  {author} {\bibinfo {author} {\bibfnamefont {A.}~\bibnamefont {Paetznick}}, \bibinfo {author} {\bibfnamefont {C.}~\bibnamefont {Knapp}}, \bibinfo {author} {\bibfnamefont {N.}~\bibnamefont {Delfosse}}, \bibinfo {author} {\bibfnamefont {B.}~\bibnamefont {Bauer}}, \bibinfo {author} {\bibfnamefont {J.}~\bibnamefont {Haah}}, \bibinfo {author} {\bibfnamefont {M.~B.}\ \bibnamefont {Hastings}},\ and\ \bibinfo {author} {\bibfnamefont {M.~P.}\ \bibnamefont {da~Silva}},\ }\bibfield  {title} {\bibinfo {title} {Performance of planar floquet codes with majorana-based qubits},\ }\href {https://doi.org/10.1103/PRXQuantum.4.010310} {\bibfield  {journal} {\bibinfo  {journal} {PRX Quantum}\ }\textbf {\bibinfo {volume} {4}},\ \bibinfo {pages} {010310} (\bibinfo {year} {2023})}\BibitemShut {NoStop}%
\bibitem [{\citenamefont {{Zhu}}\ and\ \citenamefont {{Trebst}}(2023)}]{Zhu_arxiv2023}%
  \BibitemOpen
  \bibfield  {author} {\bibinfo {author} {\bibfnamefont {G.-Y.}\ \bibnamefont {{Zhu}}}\ and\ \bibinfo {author} {\bibfnamefont {S.}~\bibnamefont {{Trebst}}},\ }\bibfield  {title} {\bibinfo {title} {{Qubit fractionalization and emergent Majorana liquid in the honeycomb Floquet code induced by coherent errors and weak measurements}},\ }\href {https://doi.org/10.48550/arXiv.2311.08450} {\bibfield  {journal} {\bibinfo  {journal} {arXiv e-prints}\ ,\ \bibinfo {eid} {arXiv:2311.08450}} (\bibinfo {year} {2023})},\ \Eprint {https://arxiv.org/abs/2311.08450} {arXiv:2311.08450 [quant-ph]} \BibitemShut {NoStop}%
\bibitem [{\citenamefont {Lavasani}\ \emph {et~al.}(2023)\citenamefont {Lavasani}, \citenamefont {Luo},\ and\ \citenamefont {Vijay}}]{Lavasani_PRB2023}%
  \BibitemOpen
  \bibfield  {author} {\bibinfo {author} {\bibfnamefont {A.}~\bibnamefont {Lavasani}}, \bibinfo {author} {\bibfnamefont {Z.-X.}\ \bibnamefont {Luo}},\ and\ \bibinfo {author} {\bibfnamefont {S.}~\bibnamefont {Vijay}},\ }\bibfield  {title} {\bibinfo {title} {Monitored quantum dynamics and the {Kitaev} spin liquid},\ }\href {https://doi.org/10.1103/PhysRevB.108.115135} {\bibfield  {journal} {\bibinfo  {journal} {Phys. Rev. B}\ }\textbf {\bibinfo {volume} {108}},\ \bibinfo {pages} {115135} (\bibinfo {year} {2023})}\BibitemShut {NoStop}%
\bibitem [{\citenamefont {Sriram}\ \emph {et~al.}(2023)\citenamefont {Sriram}, \citenamefont {Rakovszky}, \citenamefont {Khemani},\ and\ \citenamefont {Ippoliti}}]{Sriram_PRB2023}%
  \BibitemOpen
  \bibfield  {author} {\bibinfo {author} {\bibfnamefont {A.}~\bibnamefont {Sriram}}, \bibinfo {author} {\bibfnamefont {T.}~\bibnamefont {Rakovszky}}, \bibinfo {author} {\bibfnamefont {V.}~\bibnamefont {Khemani}},\ and\ \bibinfo {author} {\bibfnamefont {M.}~\bibnamefont {Ippoliti}},\ }\bibfield  {title} {\bibinfo {title} {Topology, criticality, and dynamically generated qubits in a stochastic measurement-only {Kitaev} model},\ }\href {https://doi.org/10.1103/PhysRevB.108.094304} {\bibfield  {journal} {\bibinfo  {journal} {Phys. Rev. B}\ }\textbf {\bibinfo {volume} {108}},\ \bibinfo {pages} {094304} (\bibinfo {year} {2023})}\BibitemShut {NoStop}%
\bibitem [{\citenamefont {Zhu}\ \emph {et~al.}(2024)\citenamefont {Zhu}, \citenamefont {Tantivasadakarn},\ and\ \citenamefont {Trebst}}]{Zhu_PRR2024}%
  \BibitemOpen
  \bibfield  {author} {\bibinfo {author} {\bibfnamefont {G.-Y.}\ \bibnamefont {Zhu}}, \bibinfo {author} {\bibfnamefont {N.}~\bibnamefont {Tantivasadakarn}},\ and\ \bibinfo {author} {\bibfnamefont {S.}~\bibnamefont {Trebst}},\ }\bibfield  {title} {\bibinfo {title} {Structured volume-law entanglement in an interacting, monitored majorana spin liquid},\ }\href {https://doi.org/10.1103/PhysRevResearch.6.L042063} {\bibfield  {journal} {\bibinfo  {journal} {Phys. Rev. Res.}\ }\textbf {\bibinfo {volume} {6}},\ \bibinfo {pages} {L042063} (\bibinfo {year} {2024})}\BibitemShut {NoStop}%
\bibitem [{\citenamefont {Vijayvargia}\ \emph {et~al.}(2025)\citenamefont {Vijayvargia}, \citenamefont {Day-Roberts}, \citenamefont {Botana},\ and\ \citenamefont {Erten}}]{Vijayvargia_arxiv2025}%
  \BibitemOpen
  \bibfield  {author} {\bibinfo {author} {\bibfnamefont {A.}~\bibnamefont {Vijayvargia}}, \bibinfo {author} {\bibfnamefont {E.}~\bibnamefont {Day-Roberts}}, \bibinfo {author} {\bibfnamefont {A.~S.}\ \bibnamefont {Botana}},\ and\ \bibinfo {author} {\bibfnamefont {O.}~\bibnamefont {Erten}},\ }\href {https://arxiv.org/abs/2503.09705} {\bibinfo {title} {Altermagnets with topological order in {Kitaev} bilayers}} (\bibinfo {year} {2025}),\ \Eprint {https://arxiv.org/abs/2503.09705} {arXiv:2503.09705 [cond-mat.str-el]} \BibitemShut {NoStop}%
\bibitem [{\citenamefont {Lieb}(1994)}]{Lieb_PRL1994}%
  \BibitemOpen
  \bibfield  {author} {\bibinfo {author} {\bibfnamefont {E.~H.}\ \bibnamefont {Lieb}},\ }\bibfield  {title} {\bibinfo {title} {Flux phase of the half-filled band},\ }\href {https://doi.org/10.1103/PhysRevLett.73.2158} {\bibfield  {journal} {\bibinfo  {journal} {Phys. Rev. Lett.}\ }\textbf {\bibinfo {volume} {73}},\ \bibinfo {pages} {2158} (\bibinfo {year} {1994})}\BibitemShut {NoStop}%
\bibitem [{\citenamefont {Kitaev}(2003)}]{Kitaev_AnnPhys2003}%
  \BibitemOpen
  \bibfield  {author} {\bibinfo {author} {\bibfnamefont {A.}~\bibnamefont {Kitaev}},\ }\bibfield  {title} {\bibinfo {title} {Fault-tolerant quantum computation by anyons},\ }\href {https://doi.org/https://doi.org/10.1016/S0003-4916(02)00018-0} {\bibfield  {journal} {\bibinfo  {journal} {Annals of Physics}\ }\textbf {\bibinfo {volume} {303}},\ \bibinfo {pages} {2} (\bibinfo {year} {2003})}\BibitemShut {NoStop}%
\bibitem [{\citenamefont {Rau}\ \emph {et~al.}(2014)\citenamefont {Rau}, \citenamefont {Lee},\ and\ \citenamefont {Kee}}]{Rau_PRL2014}%
  \BibitemOpen
  \bibfield  {author} {\bibinfo {author} {\bibfnamefont {J.~G.}\ \bibnamefont {Rau}}, \bibinfo {author} {\bibfnamefont {E.~K.-H.}\ \bibnamefont {Lee}},\ and\ \bibinfo {author} {\bibfnamefont {H.-Y.}\ \bibnamefont {Kee}},\ }\bibfield  {title} {\bibinfo {title} {Generic spin model for the honeycomb iridates beyond the {Kitaev} limit},\ }\href {https://doi.org/10.1103/PhysRevLett.112.077204} {\bibfield  {journal} {\bibinfo  {journal} {Phys. Rev. Lett.}\ }\textbf {\bibinfo {volume} {112}},\ \bibinfo {pages} {077204} (\bibinfo {year} {2014})}\BibitemShut {NoStop}%
\bibitem [{\citenamefont {Aaronson}\ and\ \citenamefont {Gottesman}(2004)}]{Aaronson_PRA2004}%
  \BibitemOpen
  \bibfield  {author} {\bibinfo {author} {\bibfnamefont {S.}~\bibnamefont {Aaronson}}\ and\ \bibinfo {author} {\bibfnamefont {D.}~\bibnamefont {Gottesman}},\ }\bibfield  {title} {\bibinfo {title} {Improved simulation of stabilizer circuits},\ }\href {https://doi.org/10.1103/PhysRevA.70.052328} {\bibfield  {journal} {\bibinfo  {journal} {Phys. Rev. A}\ }\textbf {\bibinfo {volume} {70}},\ \bibinfo {pages} {052328} (\bibinfo {year} {2004})}\BibitemShut {NoStop}%
\bibitem [{\citenamefont {Gottesman}(1998)}]{Gottesman_arxiv1998}%
  \BibitemOpen
  \bibfield  {author} {\bibinfo {author} {\bibfnamefont {D.}~\bibnamefont {Gottesman}},\ }\href {https://arxiv.org/abs/quant-ph/9807006} {\bibinfo {title} {The heisenberg representation of quantum computers}} (\bibinfo {year} {1998}),\ \Eprint {https://arxiv.org/abs/quant-ph/9807006} {arXiv:quant-ph/9807006 [quant-ph]} \BibitemShut {NoStop}%
\bibitem [{\citenamefont {{Vijayvargia}}\ \emph {et~al.}(2025)\citenamefont {{Vijayvargia}}, \citenamefont {{Day-Roberts}},\ and\ \citenamefont {{Erten}}}]{Vijayvargia_arxiv2025v2}%
  \BibitemOpen
  \bibfield  {author} {\bibinfo {author} {\bibfnamefont {A.}~\bibnamefont {{Vijayvargia}}}, \bibinfo {author} {\bibfnamefont {E.}~\bibnamefont {{Day-Roberts}}},\ and\ \bibinfo {author} {\bibfnamefont {O.}~\bibnamefont {{Erten}}},\ }\bibfield  {title} {\bibinfo {title} {{Error stabilized logical qubits in qudit generalizations of the monitored Kitaev model}},\ }\href {https://doi.org/10.48550/arXiv.2509.16758} {\bibfield  {journal} {\bibinfo  {journal} {arXiv e-prints}\ ,\ \bibinfo {eid} {arXiv:2509.16758}} (\bibinfo {year} {2025})},\ \Eprint {https://arxiv.org/abs/2509.16758} {arXiv:2509.16758 [quant-ph]} \BibitemShut {NoStop}%
\bibitem [{\citenamefont {Klocke}\ \emph {et~al.}(2025)\citenamefont {Klocke}, \citenamefont {Simm}, \citenamefont {Zhu}, \citenamefont {Trebst},\ and\ \citenamefont {Buchhold}}]{Klocke_PRB2025}%
  \BibitemOpen
  \bibfield  {author} {\bibinfo {author} {\bibfnamefont {K.}~\bibnamefont {Klocke}}, \bibinfo {author} {\bibfnamefont {D.}~\bibnamefont {Simm}}, \bibinfo {author} {\bibfnamefont {G.-Y.}\ \bibnamefont {Zhu}}, \bibinfo {author} {\bibfnamefont {S.}~\bibnamefont {Trebst}},\ and\ \bibinfo {author} {\bibfnamefont {M.}~\bibnamefont {Buchhold}},\ }\bibfield  {title} {\bibinfo {title} {Entanglement dynamics in monitored {Kitaev} circuits: Loop models, symmetry classification, and quantum {Lifshitz} scaling},\ }\href {https://doi.org/10.1103/PhysRevB.111.224301} {\bibfield  {journal} {\bibinfo  {journal} {Phys. Rev. B}\ }\textbf {\bibinfo {volume} {111}},\ \bibinfo {pages} {224301} (\bibinfo {year} {2025})}\BibitemShut {NoStop}%
\bibitem [{\citenamefont {Klocke}\ and\ \citenamefont {Buchhold}(2023)}]{Klocke_PRX2023}%
  \BibitemOpen
  \bibfield  {author} {\bibinfo {author} {\bibfnamefont {K.}~\bibnamefont {Klocke}}\ and\ \bibinfo {author} {\bibfnamefont {M.}~\bibnamefont {Buchhold}},\ }\bibfield  {title} {\bibinfo {title} {Majorana loop models for measurement-only quantum circuits},\ }\href {https://doi.org/10.1103/PhysRevX.13.041028} {\bibfield  {journal} {\bibinfo  {journal} {Phys. Rev. X}\ }\textbf {\bibinfo {volume} {13}},\ \bibinfo {pages} {041028} (\bibinfo {year} {2023})}\BibitemShut {NoStop}%
\bibitem [{\citenamefont {Page}(1993)}]{Page1993}%
  \BibitemOpen
  \bibfield  {author} {\bibinfo {author} {\bibfnamefont {D.~N.}\ \bibnamefont {Page}},\ }\bibfield  {title} {\bibinfo {title} {Average entropy of a subsystem},\ }\href {https://doi.org/10.1103/PhysRevLett.71.1291} {\bibfield  {journal} {\bibinfo  {journal} {Phys. Rev. Lett.}\ }\textbf {\bibinfo {volume} {71}},\ \bibinfo {pages} {1291} (\bibinfo {year} {1993})}\BibitemShut {NoStop}%
\end{thebibliography}%

\end{document}